\documentclass[fleqn,10pt]{wlscirep}
\usepackage{array}
\usepackage{makecell}
\usepackage{tabularx}
\usepackage{booktabs}
\usepackage[utf8]{inputenc}
\usepackage[T1]{fontenc}
\usepackage{algorithmic}
\usepackage{braket}
\usepackage{algorithm}
\usepackage{amsmath}
\usepackage{graphicx}
\usepackage{subcaption}
\usepackage{caption}
\usepackage{ulem}
\usepackage{xcolor}
\usepackage{multirow}
\usepackage{quantikz}
\usepackage{siunitx}  
\usepackage{booktabs} 

\usepackage{makecell}

\title{A Resource-Efficient Variational Quantum Framework for the Traveling Salesman Problem}

\author[1]{Yuefeng Lin}
\author[2]{Chao Zheng}
\author[1,*]{Cong Guo}

\affil[1]{Shenzhen SpinQ Technology Co., Ltd, Shenzhen, 518048, China}
\affil[2]{School of Energy Storage Science and Engineering, North China University of Technology, Beijing 100144, China}
\affil[*]{Correspondence: cguo@spinq.cn (C.G.)}


\begin{abstract}
The Traveling Salesman Problem (TSP) is a prototypical combinatorial
optimization problem, but its quantum implementation is limited by the
\(O(n^2)\)-qubit overhead of standard one-hot encodings. Here, we propose a
resource-efficient variational quantum framework based on compact
binary-register encoding, a permutation-preserving problem-inspired ansatz, and
a complementary divide-and-conquer execution strategy. The compact encoding
reduces the data-qubit requirement to \(O(n\log n)\), while the
divide-and-conquer formulation lowers the number of qubits required in each
local hardware execution to the size of the largest subsystem. Numerical simulations on TSP instances with 4, 5, and 6 cities achieve best
average success rates of \(100\%\), \(100\%\), and \(95.5\%\), respectively.
A local two-qubit implementation of the divide-and-conquer approximation is further evaluated for a 5-city TSP instance on SpinQ Gemini Pro and SpinQ Triangulum II NMR quantum computers.
Taken together, the results indicate how compact encoding and divide-and-conquer execution with classical post-processing can be used to study small combinatorial optimization instances on resource-constrained quantum hardware.
\end{abstract}

\begin{document}

\flushbottom
\maketitle
%
%
\thispagestyle{empty}


\section{Introduction}

Combinatorial optimization problems (COPs) arise widely in scientific computing
and engineering applications, including logistics, scheduling, network design,
and resource allocation~\cite{papadimitriou1998combinatorial}. In these
problems, the goal is to identify an optimal solution from a discrete search
space whose size typically grows exponentially with the problem dimension.
Consequently, many practically relevant COPs become computationally challenging
for exact classical algorithms as the problem scale increases.

The Traveling Salesman Problem (TSP) is one of the most extensively studied
benchmarks in combinatorial optimization. Given \(n\) cities and a pairwise
distance matrix, the objective is to find the shortest Hamiltonian cycle that
visits each city exactly once and returns to the starting city. The number of
feasible tours grows factorially with \(n\), making the TSP NP-hard and
challenging even for moderately sized instances. Classical methods for the TSP
include exact algorithms, such as the Held--Karp dynamic programming algorithm,
and heuristic or metaheuristic approaches, such as branch-and-cut,
Lin--Kernighan search, genetic algorithms, and ant colony optimization. Exact
methods provide optimality guarantees but scale exponentially in the worst case,
whereas heuristic methods often produce high-quality solutions but may require
substantial computational effort and instance-dependent parameter
tuning~\cite{held1962dynamic,applegate2011traveling}.

Quantum computing has been explored as a potential approach for tackling hard
optimization problems by exploiting quantum superposition, interference, and
entanglement. In the noisy intermediate-scale quantum (NISQ) era~\cite{preskill2018quantum}, however,
quantum processors are still limited by the number of available qubits, gate
fidelities, circuit depth, and coherence times, while fault-tolerant quantum
computation remains unavailable for many near-term optimization experiments.
Under these constraints, hybrid quantum--classical variational algorithms have
become a leading paradigm for near-term quantum optimization~\cite{cerezo2021variational,bharti2022noisy}. Representative
examples include the Variational Quantum Eigensolver (VQE), which minimizes the
expectation value of a problem Hamiltonian using a parameterized quantum
circuit optimized by a classical routine~\cite{peruzzo2014variational}, and
the Quantum Approximate Optimization Algorithm (QAOA), which prepares
approximate solutions through alternating applications of cost and mixer
operators~\cite{farhi2014quantum}.

Variational quantum algorithms have been applied to the TSP and related
routing or sequencing problems. For example, Bourreau \emph{et al.} proposed an
indirect QAOA framework for the TSP and reported results on instances with up
to eight customers~\cite{bourreau2023iqaoa}. In the context of capacitated
vehicle routing, Palackal \emph{et al.} decomposed the problem into clustering
and TSP subproblems, and investigated QAOA and VQE for the resulting TSP
instances under different hyperparameter choices~\cite{palackal2023cvrp}.
Quantum variational approaches have also been explored in computational
biology, including de novo DNA sequence reconstruction and genome assembly:
Sarkar \emph{et al.} applied QAOA to de novo DNA sequence reconstruction on a
gate-based simulator~\cite{sarkar2021quaser}, while Fang \emph{et al.}
developed a VQE-based hybrid quantum algorithm for de novo genome
assembly~\cite{PRXLife.2.023006}. These studies illustrate the broad interest
in applying NISQ algorithms to structured combinatorial optimization tasks.

Despite these developments, the efficient quantum implementation of the TSP
remains limited by the choice of problem encoding. In the standard one-hot QUBO
formulation, each city--position pair is represented by one binary variable~\cite{lucas2014ising}.
This results in \(O(n^2)\) binary variables and, under a direct qubit mapping,
\(O(n^2)\) qubits. The resulting Hamiltonian also contains constraint terms
that penalize invalid assignments, increasing the complexity of the
optimization landscape. Such resource requirements are poorly matched to
current NISQ hardware. To address this bottleneck, compact encodings have been
investigated. In particular, Ramezani \emph{et al.} proposed a higher-order
binary formulation that reduces the qubit requirement for the TSP from
\(O(n^2)\) to \(O(n\log n)\), improving qubit efficiency at the encoding
level~\cite{ramezani2024reducing}. In parallel, distributed and decomposed
variational frameworks have been proposed as a means of extending quantum
optimization beyond the capacity of a single small processor by partitioning a
problem into smaller quantum-executable components~\cite{khait2023distributed}.

Here we develop a resource-efficient variational framework for the TSP that targets two distinct near-term hardware regimes. For processors that can support compact multi-register circuits, we combine cyclic-symmetry reduction with a binary city-label encoding, so that the \(M=n-1\) unfixed tour positions require \(M\lceil\log_2 M\rceil\) data qubits. We then construct the corresponding Hamiltonian in terms of distance, repeated-city, and invalid-code projectors. On this representation, a permutation-preserving ansatz uses parameterized register-wise SWAP operations to move amplitude among valid tours while keeping the data registers in the feasible permutation subspace.

For devices with more severe qubit limits, we introduce a complementary divide-and-conquer execution strategy. Rather than preparing the full binary-register state, this strategy factorizes the global variational state into smaller subsystems and reconstructs the global Hamiltonian expectation value from local measurements. This reduces the qubit count required in each quantum execution to the size of the largest subsystem, while making explicit the accompanying product-state approximation and classical reconstruction cost. Because this strategy does not preserve global TSP feasibility at the state-preparation level, the repetition and invalid-code penalties are retained in the objective.

We evaluate these two components in their respective operating regimes. The permutation-preserving ansatz is tested in simulations on 4--6 city TSP instances, where it achieves best average success rates of \(100\%\), \(100\%\), and \(95.5\%\), respectively, and improves over a prior one-hot feasible-subspace ansatz under the reported benchmark protocol. The divide-and-conquer strategy is implemented on SpinQ Gemini Pro and SpinQ Triangulum II NMR quantum computers as a two-qubit local-execution demonstration for a 5-city instance.

The central contribution of this work is a resource-aware variational framework that connects a compact TSP encoding, a feasible-subspace search circuit, and a local-execution strategy for devices with very limited qubit counts. Specifically,
\begin{itemize}
    \item we formulate a symmetry-reduced binary-register Hamiltonian for the TSP that uses register-wise city-label projectors to express the distance, repeated-city, and invalid-code terms in a compact \(M\lceil\log_2 M\rceil\)-qubit representation;

    \item we design a permutation-preserving ansatz that mixes feasible tours by parameterized register-wise SWAP operations, thereby replacing one-hot permutation-matrix dynamics with a binary-register mixing mechanism over the feasible tour space;

    \item we introduce a complementary divide-and-conquer execution strategy for highly resource-constrained hardware, in which local subsystem measurements are classically recombined to estimate the diagonal binary-register Hamiltonian while retaining penalty terms for global feasibility;

    \item we validate the two resource regimes separately: the binary-register feasible-subspace ansatz is benchmarked in simulations on 4--6 city instances, whereas the divide-and-conquer strategy is demonstrated as a two-qubit local execution of a 5-city instance on SpinQ NMR quantum computers.

\end{itemize}

The remainder of this paper is organized as follows. Section~2 presents the
methodology, including the one-hot QUBO formulation, the compact binary-register
Hamiltonian with symmetry reduction, the permutation-preserving ansatz, and the
divide-and-conquer variational framework. Section~3 reports the simulation and
hardware results, including the depth-dependent performance of the proposed
ansatz, comparison with a prior feasible-subspace ansatz, and experimental
validation on real quantum hardware. Section~4 concludes the paper and
discusses future research directions.

\section{Methodology}
\subsection{One-hot QUBO formulation}

We first recall the standard one-hot QUBO formulation of the Traveling Salesman
Problem (TSP), which serves as a conventional baseline representation. For an
\(n\)-city TSP, let \(x_{i,u}\in\{0,1\}\) indicate whether city \(u\) is visited
at tour position \(i\). A valid tour must satisfy two sets of assignment
constraints,
\begin{equation}
\sum_{u=0}^{n-1}x_{i,u}=1,\quad \forall i,
\qquad
\sum_{i=0}^{n-1}x_{i,u}=1,\quad \forall u.
\label{eq:onehot_constraints}
\end{equation}
The first constraint ensures that each position is occupied by exactly one
city, whereas the second ensures that each city appears exactly once.

The corresponding QUBO objective is given by
\begin{equation}
C_{\mathrm{QUBO}}=
\sum_{i=0}^{n-1}\sum_{u=0}^{n-1}\sum_{v=0}^{n-1}
D_{uv}\,x_{i,u}x_{i+1,v}
+
A\sum_{i=0}^{n-1}
\left(1-\sum_{u=0}^{n-1}x_{i,u}\right)^2
+
B\sum_{u=0}^{n-1}
\left(1-\sum_{i=0}^{n-1}x_{i,u}\right)^2,
\label{eq:qubo_objective}
\end{equation}
where \(D_{uv}\) denotes the distance between cities \(u\) and \(v\), and
\(A\) and \(B\) are positive penalty coefficients. The index \(i+1\) is taken
modulo \(n\), such that the last city is connected back to the first city.

Using the standard mapping from binary variables to Pauli operators,
\begin{equation}
x_{i,u}\mapsto \frac{1-Z_{i,u}}{2},
\label{eq:qubo_to_ising}
\end{equation}
the QUBO objective can be converted into an Ising Hamiltonian,
\begin{equation}
H_{\mathrm{QUBO}}
=
H_{\mathrm{dist}}
+
A H_{\mathrm{pos}}
+
B H_{\mathrm{city}}.
\label{eq:onehot_hamiltonian}
\end{equation}
Although this formulation is straightforward, it assigns one binary variable
to each city--position pair and therefore requires \(n^2\) qubits. This
quadratic qubit overhead limits its applicability on near-term quantum
hardware.

\subsection{Binary-register Hamiltonian with symmetry reduction}

To reduce the qubit requirement, we encode city labels directly using binary
registers. We also exploit the cyclic symmetry of TSP tours: since cyclic
shifts of the same Hamiltonian cycle correspond to the same tour and have the
same total length, one city can be fixed as the starting city without loss of
generality. This symmetry reduction removes redundant representations and
leaves only the remaining
\begin{equation}
M=n-1
\end{equation}
cities to be ordered.

Each of the \(M\) remaining tour positions is represented by a \(k\)-qubit
binary register, where
\begin{equation}
k=\lceil \log_2 M\rceil .
\label{eq:k_binary_register}
\end{equation}
Thus, the compact representation requires
\(M\lceil\log_2 M\rceil\) data qubits after symmetry reduction. 

For the \(i\)-th register, the projector onto the binary code
\(a\in\{0,\dots,2^k-1\}\) is defined as
\begin{equation}
P_i(a)
=
\prod_{b=0}^{k-1}
\frac{1+(-1)^{a_b}Z_{i,b}}{2},
\label{eq:binary_projector}
\end{equation}
where \(a_b\) is the \(b\)-th bit of \(a\), and \(Z_{i,b}\) denotes the
Pauli-\(Z\) operator acting on the \(b\)-th qubit of register \(i\). The codes
\(0,\dots,M-1\) represent valid reduced city labels, while the remaining codes
\(M,\dots,2^k-1\) are invalid.

The Hamiltonian consists of three components. The first term penalizes repeated
valid city labels:
\begin{equation}
H_{\mathrm{rep}}
=
\sum_{i=0}^{M-2}
\sum_{j=i+1}^{M-1}
\sum_{a=0}^{M-1}
P_i(a)P_j(a).
\label{eq:repetition_penalty}
\end{equation}
The second term penalizes invalid binary codes:
\begin{equation}
H_{\mathrm{inv}}
=
\sum_{i=0}^{M-1}
\sum_{a=M}^{2^k-1}
P_i(a).
\label{eq:invalid_penalty}
\end{equation}
The third term evaluates the tour length. Let \(s\) denote the fixed starting
city. The reduced city labels \(a,b\in\{0,\dots,M-1\}\) correspond to the
remaining original cities, and their pairwise distances are denoted by
\(\tilde{D}_{ab}\). The distance Hamiltonian is
\begin{equation}
H_{\mathrm{dist}}
=
\sum_{a=0}^{M-1}
D_{s,a}P_0(a)
+
\sum_{i=0}^{M-2}
\sum_{a=0}^{M-1}
\sum_{b=0}^{M-1}
\tilde{D}_{ab}P_i(a)P_{i+1}(b)
+
\sum_{a=0}^{M-1}
D_{a,s}P_{M-1}(a).
\label{eq:reduced_distance_hamiltonian}
\end{equation}
The complete binary-register Hamiltonian is then
\begin{equation}
H
=
H_{\mathrm{dist}}
+
\lambda H_{\mathrm{rep}}
+
\mu H_{\mathrm{inv}},
\label{eq:binary_hamiltonian}
\end{equation}
where \(\lambda\) and \(\mu\) are positive penalty coefficients. 

\subsection{Permutation-preserving problem-inspired ansatz}
\label{pia}
The elementary block of the ansatz is an ancilla-assisted parameterized
register-swap operation between two neighboring city registers. Let
\(q_{\mathrm{aux}}\) denote the auxiliary qubit, and let
\(S_{i,i+1}^{(\mathrm{aux})}\) be the controlled register-swap operator between
registers \(i\) and \(i+1\), defined as
\begin{equation}
S_{i,i+1}^{(\mathrm{aux})}
=
\prod_{b=0}^{k-1}
\mathrm{CSWAP}_{q_{\mathrm{aux}},\,i_b,\,(i+1)_b},
\label{eq:controlled_register_swap}
\end{equation}
where \(i_b\) and \((i+1)_b\) denote the \(b\)-th qubits of the two registers.
The parameterized register-swap block is implemented as
\begin{equation}
U_{\mathrm{swap}}^{(i,i+1)}(\theta)
=
R_y^{(q_{\mathrm{aux}})}(2\theta)\,
S_{i,i+1}^{(\mathrm{aux})}\,.
\label{eq:parameterized_register_swap}
\end{equation}
Thus, the variational parameter is introduced through rotations on the auxiliary
qubit, which coherently controls the exchange of two complete binary city
registers. The auxiliary qubit is reused throughout the circuit, and final tour
probabilities are obtained from the marginal distribution of the data registers.

One ansatz layer is defined as an ordered product of neighboring register-swap
blocks,
\begin{equation}
U_{\mathrm{layer}}(\boldsymbol{\theta}^{(\ell)})
=
\overrightarrow{\prod_{i=0}^{M-2}}
U_{\mathrm{swap}}^{(i,i+1)}
\left(\theta_{\ell,i}\right),
\label{eq:permutation_layer}
\end{equation}
where the arrow denotes ascending order of \(i\), and
\(\boldsymbol{\theta}^{(\ell)}
=
\{\theta_{\ell,0},\dots,\theta_{\ell,M-2}\}\).
Starting from the auxiliary state \(|0\rangle_{\mathrm{aux}}\) and the canonical
feasible permutation \(|\pi_0\rangle\), the full variational state after \(L\)
layers is
\begin{equation}
|\Psi(\boldsymbol{\theta})\rangle
=
\prod_{\ell=1}^{L}
U_{\mathrm{layer}}(\boldsymbol{\theta}^{(\ell)})
\left(|0\rangle_{\mathrm{aux}}\otimes|\pi_0\rangle\right).
\label{eq:permutation_ansatz_state}
\end{equation}
The number of trainable parameters is
\begin{equation}
N_{\mathrm{param}}
=
L(M-1)
=
L(n-2).
\label{eq:number_parameters}
\end{equation}

Since adjacent transpositions generate the symmetric group, repeated application
of neighboring register-swap blocks can connect the initial ordering to any
feasible permutation. Because the auxiliary qubit may remain entangled with the
data registers, the probability assigned to a feasible tour \(\pi_k\) is defined
as the marginal probability over the auxiliary qubit,
\begin{equation}
P_k
=
\sum_{\alpha\in\{0,1\}}
\left|
\langle \alpha|_{\mathrm{aux}}
\langle \pi_k|
\Psi(\boldsymbol{\theta})\rangle
\right|^2 .
\label{eq:measurement_probability}
\end{equation}

The ansatz preserves feasibility on the data registers by construction. Each
controlled register-swap operation either leaves two complete city registers
unchanged or exchanges them as whole binary labels. Therefore, starting from a
valid permutation, the data-register state remains within the feasible
permutation subspace and cannot acquire support on repeated-city or invalid-code
configurations. Consequently, the repetition and invalid-code penalty terms in
Eq.~\eqref{eq:binary_hamiltonian} can be omitted, and the variational objective
is evaluated using only the distance Hamiltonian,
\begin{equation}
E(\boldsymbol{\theta})
=
\langle \Psi(\boldsymbol{\theta})|
\left(I_{\mathrm{aux}}\otimes H_{\mathrm{dist}}\right)
|\Psi(\boldsymbol{\theta})\rangle .
\label{eq:permutation_ansatz_energy}
\end{equation}

The total qubit requirement is therefore
\(M\lceil\log_2 M\rceil+1\), including one coherently reused auxiliary qubit.
Although this ansatz is more compact than the one-hot formulation, it may still
be too large for extremely constrained quantum processors. For example, for an
\(n=6\) TSP instance, symmetry reduction gives \(M=5\) and
\(k=\lceil\log_2 5\rceil=3\), resulting in 16 qubits including
the ancilla. This exceeds the capacity of the two-qubit NMR processor used in
our hardware experiments. We therefore introduce, in the following subsection,
a complementary divide-and-conquer framework targeted at more restrictive
hardware settings. The two approaches address different resource regimes and
are not combined within a single experiment.

\subsection{Divide-and-Conquer Quantum Optimization Framework}
While the problem-inspired ansatz of Section~\ref{pia} already operates
within the compact \(O(n\log n)\)-qubit space, even this reduced
requirement can exceed the capacity of extremely resource-constrained
quantum hardware such as the two-qubit NMR processors considered in
our experiments. We therefore introduce, as an alternative
resource-reduction strategy, a divide-and-conquer framework that
decouples the global \(n\)-register state into a tensor product of
smaller subsystem states. Unlike the problem-inspired ansatz, which
employs entangling SWAP operations across different city
registers to enforce the permutation structure, the divide-and-conquer
framework treats each subsystem independently at the state-preparation
level and recombines local expectation values classically.
The key observation is that the Hamiltonian derived in the previous subsection is diagonal in the computational basis and can be written as a linear combination of tensor-product operators,
\begin{equation}
H = \sum_t c_t H_t,
\end{equation}
where \(c_t \in \mathbb{R}\), and each term takes the form
\begin{equation}
H_t = \bigotimes_{i=1}^{N} \tilde{O}_{t,i}, \qquad \tilde{O}_{t,i} \in \{I,Z\}^{\otimes q_i}.
\end{equation}
Here, \(N\) denotes the number of logical subsystems, and \(\tilde{O}_{t,i}\) is the local operator acting on $q_i$-qubits subsystem \(i\). Since all terms consist only of \(I\) and \(Z\) operators, they mutually commute and are directly measurable in the computational basis.

Assume that the variational state is prepared as a product state over subsystems,
\begin{equation}
|\psi(\boldsymbol{\theta})\rangle = \bigotimes_{i=1}^{N} |\psi_i(\boldsymbol{\theta}_i)\rangle.
\label{eq:product_state}
\end{equation}
Then, for each Hamiltonian term \(H_t\), the expectation value factorizes as
\begin{equation}
\langle \psi(\boldsymbol{\theta})| H_t |\psi(\boldsymbol{\theta})\rangle
=
\prod_{i=1}^{N}
\langle \psi_i(\boldsymbol{\theta}_i)| \tilde{O}_{t,i} |\psi_i(\boldsymbol{\theta}_i)\rangle.
\end{equation}
Accordingly, the global variational objective can be evaluated as
\begin{equation}
E(\boldsymbol{\theta})
=
\langle \psi(\boldsymbol{\theta})|H|\psi(\boldsymbol{\theta})\rangle
=
\sum_t c_t
\prod_{i=1}^{N}
\langle \psi_i(\boldsymbol{\theta}_i)| \tilde{O}_{t,i} |\psi_i(\boldsymbol{\theta}_i)\rangle.
\label{eq:factorized_energy}
\end{equation}
Equation~\eqref{eq:factorized_energy} permits the global energy to be reconstructed from expectation values measured on smaller subsystems. The reduction in simultaneous qubit count is accompanied by problem-dependent Hamiltonian-term counts, local measurement requirements, and classical post-processing. In the present implementation, this formulation is used to execute a symmetry-reduced 5-city instance on two-qubit NMR hardware.

 The product-state assumption in Eq.~\eqref{eq:product_state} applies only to the factorization between subsystems. Each local state \(|\psi_i(\boldsymbol{\theta}_i)\rangle\) may still be prepared by a parameterized circuit containing entangling gates, such as CZ or CNOT operations, within subsystem \(i\). The hardware-efficient circuit in Fig.~\ref{fig:hardware_ansatz} is one such two-qubit local ansatz. Thus, the approximation removes inter-subsystem entanglement while retaining intra-subsystem variational structure. As a consequence, global feasibility constraints are not enforced by state preparation, in contrast to the permutation-preserving ansatz of Section~2.3. Therefore, the repetition and invalid-code penalty
terms in Eq.~\eqref{eq:binary_hamiltonian} must be retained in the Hamiltonian for the divide-and-conquer implementation. 

\section{Results}

\subsection{Experimental Setup}
The proposed framework is evaluated using numerical simulations and hardware experiments. The simulations examine the permutation-preserving ansatz in the compact binary encoding, whereas the hardware experiments assess the divide-and-conquer execution strategy on resource-constrained NMR devices.

\paragraph{Simulation setup.}
We consider TSP instances with \(n=4,5,6\) cities. After applying symmetry reduction by fixing the starting city, each instance is encoded with \(M=n-1\) city registers and \(k=\lceil \log_2 M\rceil\) qubits per register. For each problem size, we generate 10 fully connected graphs with symmetric edge weights drawn uniformly from \([10,50]\). For each graph instance, the variational circuit is optimized from 100 random initializations, and the circuit depth \(L\) is swept from \(n-1\) to 30 layers. Thus, for each pair \((n,L)\), we perform 1000 optimization runs in total.

The random graph instances are solved exactly by exhaustive enumeration to identify the global optimum before the variational runs are evaluated. To make the benchmark reproducible, the distance matrices, random seeds, optimal tours, and optimal tour lengths for all generated instances are listed in Supplementary Tables~S1--S3.

The success rate is defined as the fraction of runs that recover the optimal
TSP tour. Specifically, a run is counted as successful if the feasible tour with the highest final measurement probability belongs to the set of globally optimal tours. For each graph instance, we compute the success rate over 100 random
initializations, and then report the average success rate over the 10 graph
instances. The shaded regions in the figures indicate the absolute performance
range across the 10 graph instances, bounded by the minimum and maximum success
rates. All simulations are implemented in the SpinQit framework~\cite{SpinQit}
using the Torch simulator, and circuit parameters are optimized with Adam.

\begin{figure}[ht]
    \centering
    \[
    \begin{quantikz}
    \lstick{$q_0$} & \gate{R_y(\theta_1)} & \ctrl{1} & \gate{R_y(\theta_3)} & \qw \\
    \lstick{$q_1$} & \gate{R_y(\theta_2)} & \gate{Z}  & \gate{R_y(\theta_4)} & \qw
    \end{quantikz}
    \]
    \caption{Two-qubit variational circuit used in the hardware experiments.}
    \label{fig:hardware_ansatz}
\end{figure}
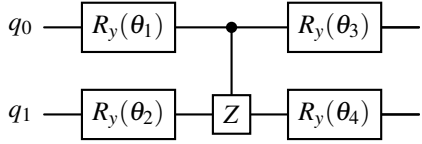

\paragraph{Hardware setup.}
Hardware experiments are conducted on the SpinQ Gemini Pro and Triangulum II NMR quantum computers~\cite{hou2021spinq,feng2022spinq}. The divide-and-conquer framework is applied to a representative 5-city TSP instance using two physical qubits per subsystem. After symmetry reduction, the instance corresponds to an 8-qubit problem Hamiltonian. The global problem is partitioned into four logical subsystems, each represented by a 2-qubit variational state prepared by the hardware-efficient ansatz shown in Fig.~\ref{fig:hardware_ansatz}. The subsystem expectation values are measured on the hardware and combined classically to reconstruct the total energy. The distance matrix, exact optimal tour, optimal tour length are reported in the Supplementary Information.

The hardware circuit parameters are optimized using SPSA for
\(N_{\mathrm{iter}} = 120\)  iterations. At each iteration, the expectation value of
the factorized objective in Eq.~\eqref{eq:factorized_energy} is estimated by
measuring all required local computational-basis observables. Each circuit is
executed with \(N_{\mathrm{shots}} = 1024\) shots. To assess the influence of hardware
noise and measurement errors, we compare four optimization trajectories: raw
hardware measurements, iterative Bayesian unfolding (IBU) measurement-error
mitigation, matrix-inversion-based measurement-error mitigation, and the
noise-free simulator.

For measurement-error mitigation, a two-qubit readout calibration matrix is
obtained immediately before each hardware experiment. Specifically, the four
computational basis states \(|00\rangle\), \(|01\rangle\), \(|10\rangle\), and
\(|11\rangle\) are prepared on the device, and their measured output
distributions are used to construct the calibration matrix.

\subsection{Simulation performance of the problem-inspired ansatz}

\begin{figure}[htbp]
    \centering
    \includegraphics[width=0.6\textwidth]{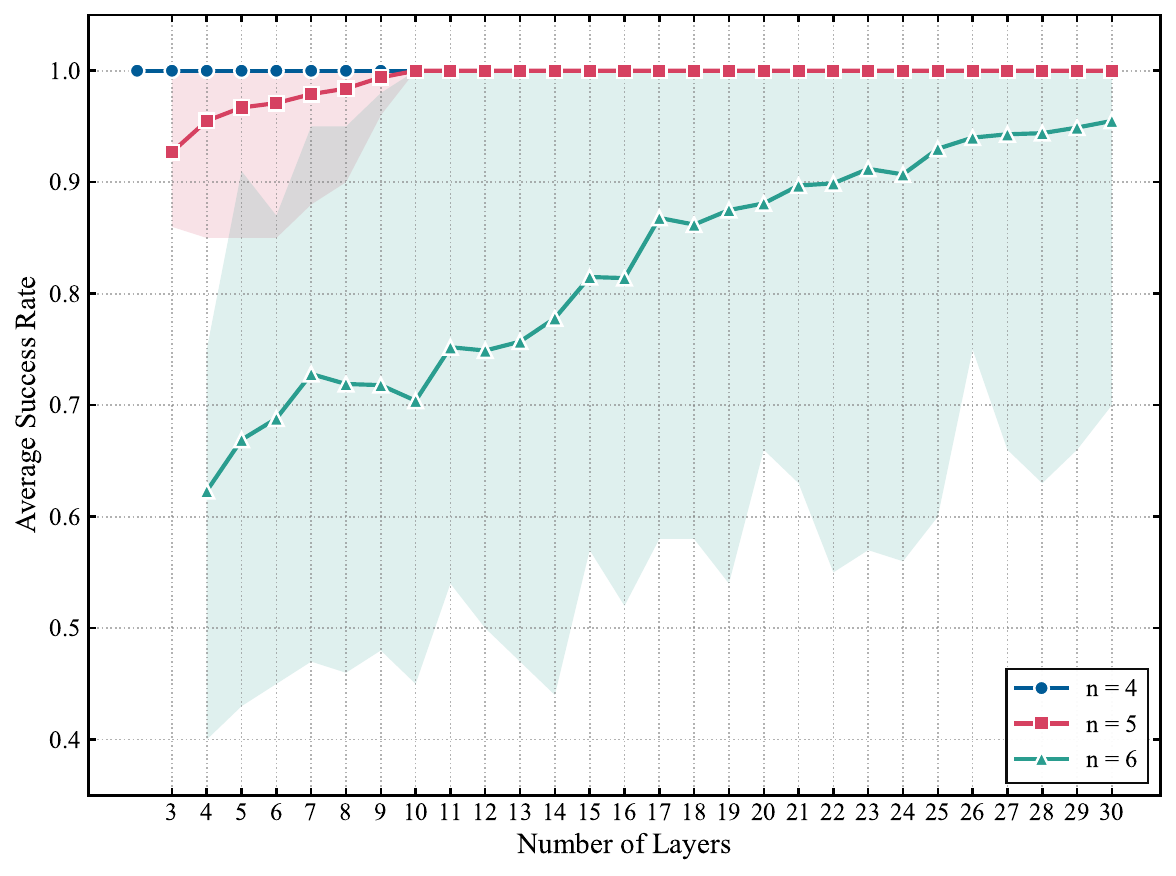}
    \caption{Success rate as a function of circuit depth for the proposed problem-inspired ansatz. Results are shown for TSP instances with \(n=4,5,6\) cities. Solid curves denote the average success rate over 10 randomly generated graph instances. Shaded regions indicate the absolute performance range across the 10 instances, bounded by the minimum and maximum success rates.}
    \label{fig:success_rate_vs_layers}
\end{figure}

We assessed the dependence of the proposed problem-inspired ansatz on circuit depth for TSP instances with \(n=4,5,6\) cities. Figure~\ref{fig:success_rate_vs_layers} reports the average success rate as a function of the ansatz depth \(L\). The solid curves denote averages over 10 independently generated graph instances, whereas the shaded regions show the range between the minimum and maximum success rates across these instances.

The results show a depth dependence that becomes more pronounced as the problem size increases. For the 4-city instances, the success rate is \(100\%\) for all tested depths, with no variation across graph instances. This behavior is consistent with the small feasible solution space obtained after symmetry reduction.

For the 5-city instances, the average success rate is \(92.7\%\) at the smallest tested depth and increases to \(99.4\%\) at \(L=9\). From \(L=10\) onward, the average success rate reaches \(100\%\) for the sampled instances. The narrowing of the shaded region indicates reduced instance-to-instance variation at larger depths.

The 6-city instances exhibit a more gradual dependence on circuit depth. The average success rate increases from \(62.3\%\) at the smallest tested depth to \(95.5\%\) at the largest tested depth, with minor non-monotonic variations at intermediate depths. The maximum success rate reaches \(100\%\) for some depths, whereas the minimum success rate increases from \(40.0\%\) to \(70.0\%\), indicating improved lower-tail performance across the sampled instances.

These observations suggest that additional register-swap layers improve the ability of the ansatz to redistribute probability within the feasible permutation subspace, particularly for the larger instances considered here. They also indicate a depth--performance trade-off: shallow circuits are adequate for the smallest instances, whereas larger instances require deeper circuits to obtain consistently high success rates under the present simulation protocol.

\subsection{Comparison with a prior feasible-subspace ansatz}

We next compare the best-achieved performance of the proposed ansatz with the problem-inspired feasible-subspace ansatz reported in Ref.~\cite{matsuo2023enhancing}. That ansatz uses a one-hot TSP encoding and constructs variational circuits that remain within the feasible tour subspace. Because both methods enforce feasibility at the circuit level, the comparison focuses on the encoding and parameterization of the feasible subspace rather than on constraint satisfaction alone.

The comparison is conducted for TSP instances with \(n=4,5,6\) cities. For the proposed method, we use the best average success rate obtained over the tested circuit depths. For each problem size, the reported value corresponds to the average success rate over 10 randomly generated graph instances, and the corresponding range is bounded by the minimum and maximum success rates across these instances. This comparison should be interpreted as a performance and resource-footprint comparison under the reported experimental settings, rather than as a fully gate-budget-matched benchmark, because the two ansatz families use different encodings and different native circuit primitives.

\begin{figure}[htbp]
    \centering
    \includegraphics[width=0.8\textwidth]{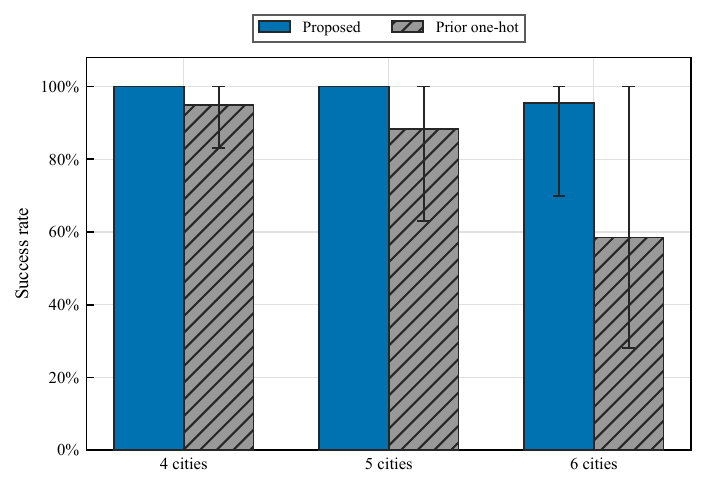}
    \caption{Performance comparison between the proposed problem-inspired ansatz and the prior one-hot feasible-subspace ansatz reported in Ref.~\cite{matsuo2023enhancing}. Bars denote the average success rate over 10 randomly generated TSP instances for each problem size. Error bars indicate the absolute performance range across instances, bounded by the minimum and maximum success rates.}
    \label{fig:algorithm_comparison}
\end{figure}

As shown in Fig.~\ref{fig:algorithm_comparison}, the proposed ansatz yields higher average success rates than the prior one-hot feasible-subspace construction for the tested instances. For the 4-city instances, the proposed method reaches an average success rate of \(100\%\), compared with \(94.9\%\) for the prior ansatz, and the minimum success rate increases from \(83.0\%\) to \(100\%\). For the 5-city instances, the proposed method also reaches an average success rate of \(100\%\), whereas the prior method reports \(88.3\%\); the minimum success rate increases from \(63.0\%\) to \(100\%\).

The difference is largest for the 6-city instances. The proposed method achieves an average success rate of \(95.5\%\), whereas the prior feasible-subspace ansatz attains \(58.5\%\), corresponding to an absolute difference of \(37.0\) percentage points. The minimum success rate increases from \(28.0\%\) to \(70.0\%\), although both methods reach a maximum success rate of \(100\%\) on at least one graph instance. This pattern indicates that the one-hot ansatz can solve favorable instances, while the binary-register ansatz shows less variation across the sampled instances under the reported protocol.

Because both methods remain within the feasible TSP subspace, the observed differences are attributable to the encoding strategy and feasible-state parameterization rather than feasibility preservation alone. The prior method uses a one-hot representation, requiring \(M^2\) qubits for the reduced \(M\)-city TSP representation and representing tours as permutation matrices. The proposed method instead uses compact binary city registers and generates valid permutations through controlled register swaps. This construction reduces the embedding overhead and defines a structured, layer-wise mixing process over feasible permutations.

The resource comparison in Table~\ref{tab:resource_compare} further highlights the structural differences between the two feasible-subspace ansatze. For an \(M\)-city reduced TSP representation, the proposed method requires \(M\lceil \log_2 M\rceil+1\) qubits, whereas the prior construction requires \(M^2\) qubits. Thus, the proposed encoding reduces the qubit requirement from quadratic to near-linear scaling with respect to \(M\). The number of trainable parameters in the proposed circuit is \((M-1)L\), where \(L\) is the number of variational layers, while the prior method uses \(\frac{1}{2}M^2-\frac{1}{2}M\) parameters. This makes the expressibility of the proposed circuit tunable through the circuit depth. Although the proposed construction uses \(\lceil \log_2 M\rceil(M-1)L\) CSWAP gates, and each CSWAP must be decomposed into hardware-native gates on most devices, the reduced qubit footprint and depth-controllable structure provide a compact representation for the problem sizes considered.

Overall, the comparison suggests that the performance differences observed in the tested small-scale instances arise from the compact binary-register representation and its feasible-space mixing mechanism, in addition to feasibility preservation. These results motivate further matched-budget benchmarks on larger instances and hardware-native decompositions.

\begin{table}[htbp]
\centering
\caption{Resource complexity comparison between the proposed permutation-subspace parameterized quantum circuit and the prior construction for an \(M\)-city reduced TSP representation. Here, \(L\) is the number of variational layers, and \(\operatorname{popcount}(i)\) denotes the Hamming weight of integer \(i\). The prior-work formulas are taken from Table~I of Ref.~\cite{matsuo2023enhancing}, where they are originally written in terms of \(n\); we rewrite them in terms of \(M\) for consistency with the symmetry-reduced notation used here.}
\label{tab:resource_compare}
\begin{tabular}{lcc}
\toprule
\textbf{Resource type} & \textbf{Proposed method} & \textbf{Prior work} \\
\midrule
Number of qubits       & $M\lceil \log_2 M \rceil + 1$                     & $M^2$                  \\
Number of parameters       & $(M-1)L$                     & $\frac{1}{2}M^2 - \frac{1}{2}M$                  \\
Number of one-qubit gates  & $(M-1)L + \sum_{i=0}^{M-1} \operatorname{popcount}(i)$                     & $M^2 - 1$             \\
Number of two-qubit gates  & --                    & $M^2 - M + 2$              \\
Number of CSWAP gates      & $\lceil \log_2 M \rceil\,(M-1)L$                    & $\frac{1}{3}M^3 - \frac{1}{2}M^2 + \frac{1}{6}M - 1$              \\
\bottomrule
\end{tabular}
\end{table}

\subsection{Experimental Validation on Real Quantum Hardware}

\begin{figure}[ht]
    \centering
    \begin{subfigure}[b]{0.48\textwidth}
        \centering
        \includegraphics[width=\textwidth]{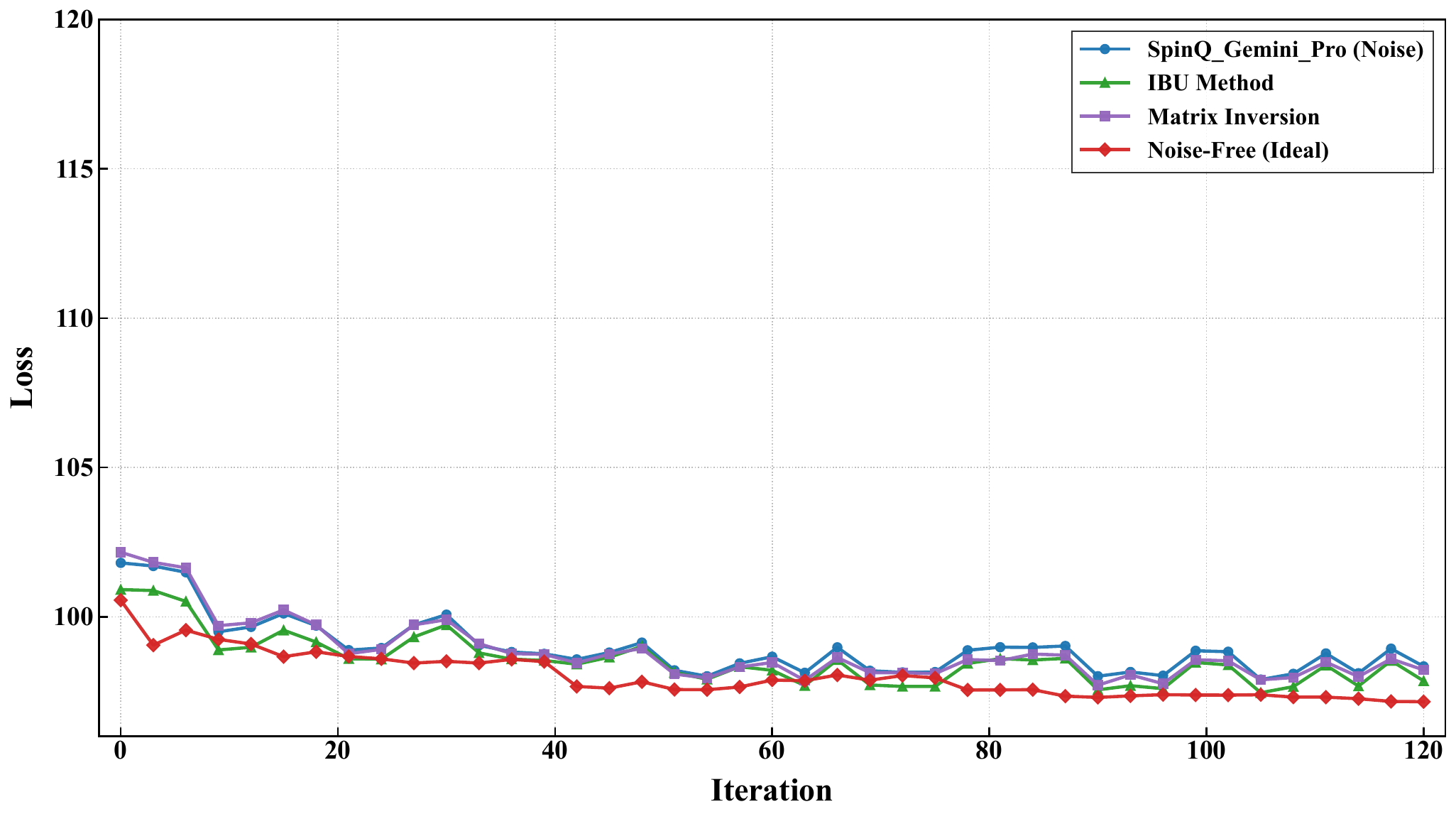}
        \caption{Loss on SpinQ Gemini Pro.}
        \label{fig:gemini_loss}
    \end{subfigure}
    \hfill
    \begin{subfigure}[b]{0.48\textwidth}
        \centering
        \includegraphics[width=\textwidth]{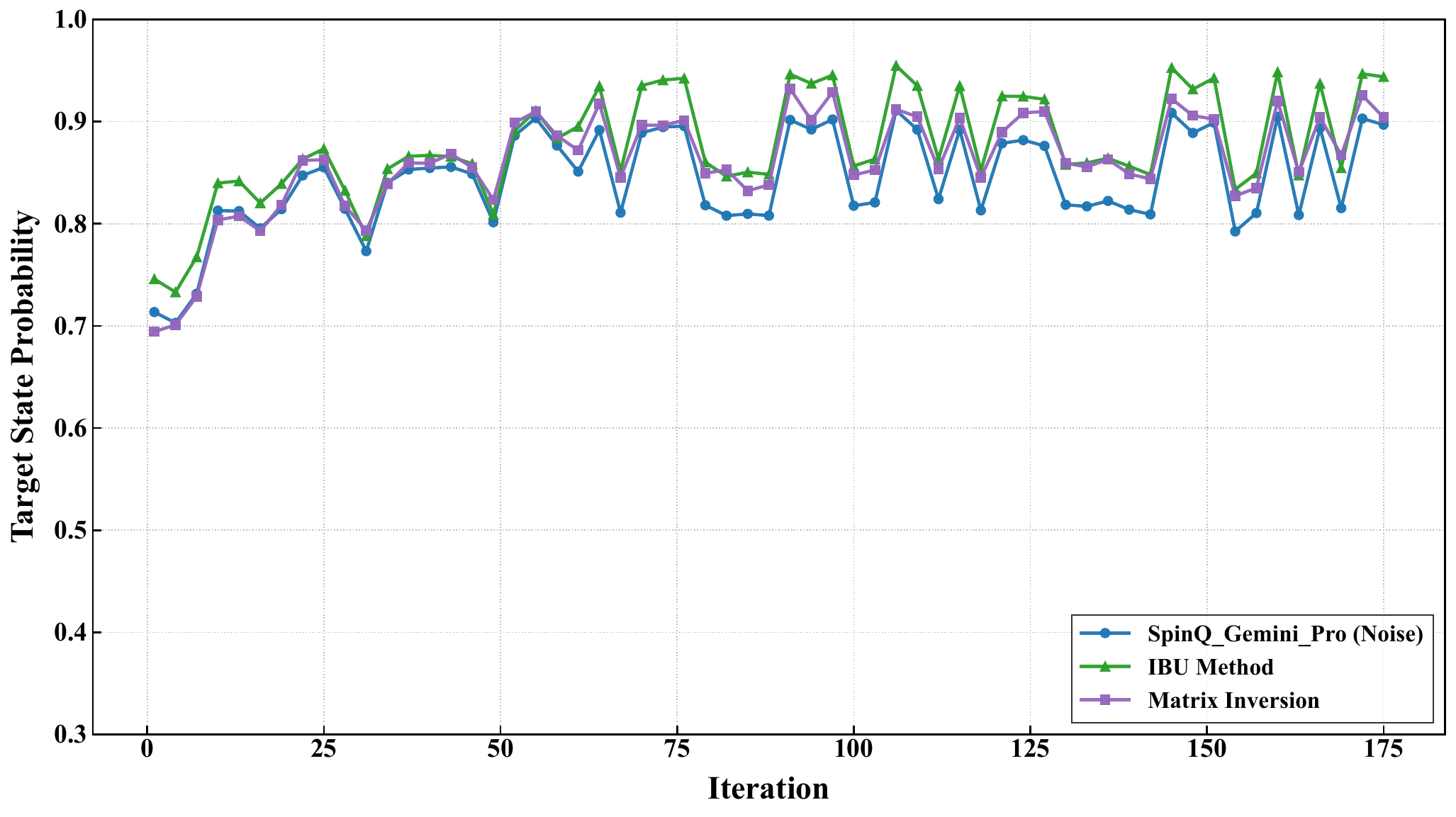}
        \caption{Target-state probability on SpinQ Gemini Pro.}
        \label{fig:gemini_prob}
    \end{subfigure}
    \caption{Optimization trajectories on SpinQ Gemini Pro for the 5-city TSP instance. Panel (a) shows the loss values obtained from raw noisy measurements, iterative Bayesian unfolding (IBU), matrix-inversion-based mitigation, and the noise-free simulator. Panel (b) shows the corresponding target-state probabilities during optimization.}
    \label{fig:gemini_hardware}
\end{figure}

\begin{figure}[ht]
    \centering
    \begin{subfigure}[b]{0.48\textwidth}
        \centering
        \includegraphics[width=\textwidth]{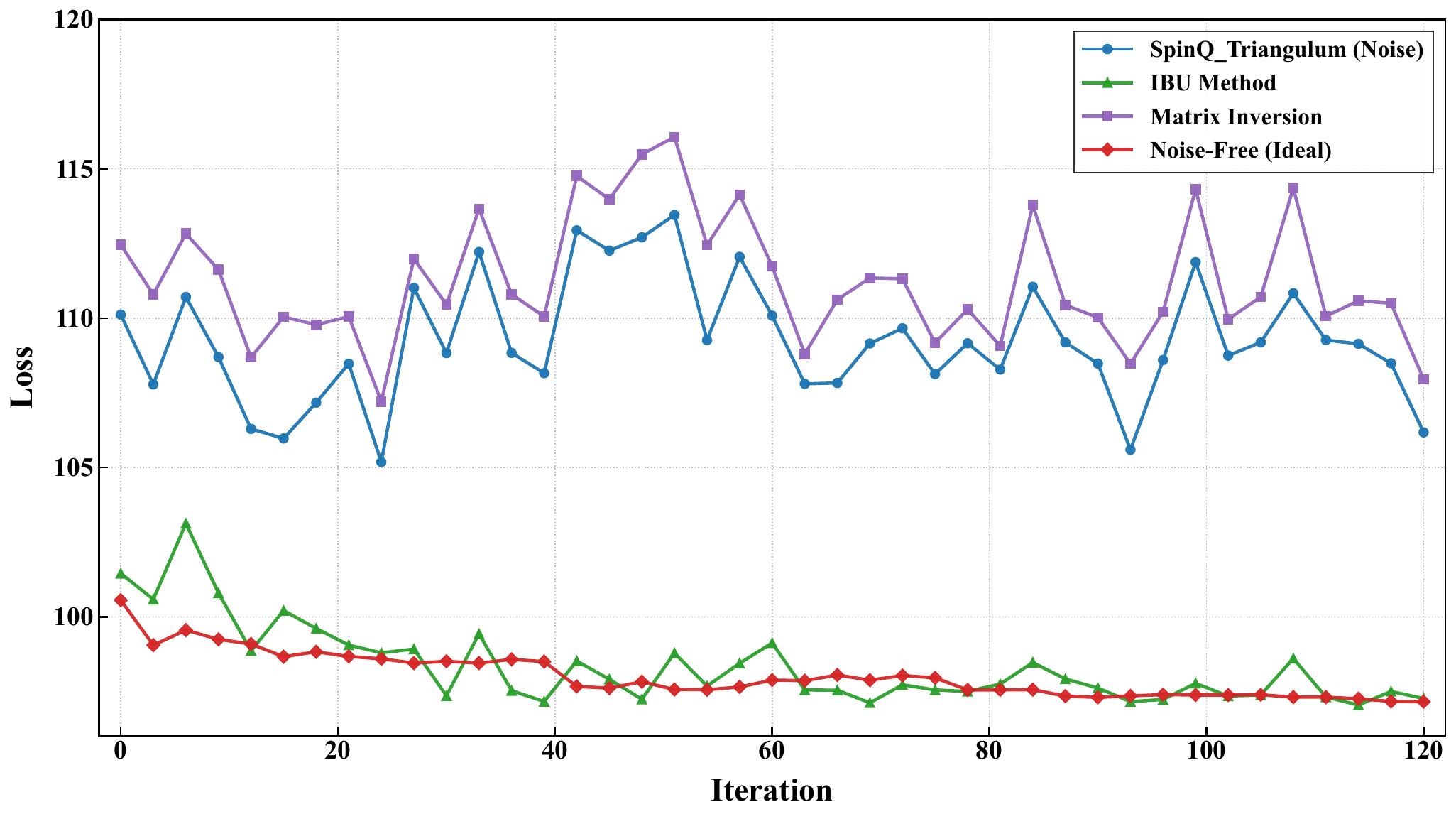}
        \caption{Loss on Triangulum II.}
        \label{fig:triangulum_loss}
    \end{subfigure}
    \hfill
    \begin{subfigure}[b]{0.48\textwidth}
        \centering
        \includegraphics[width=\textwidth]{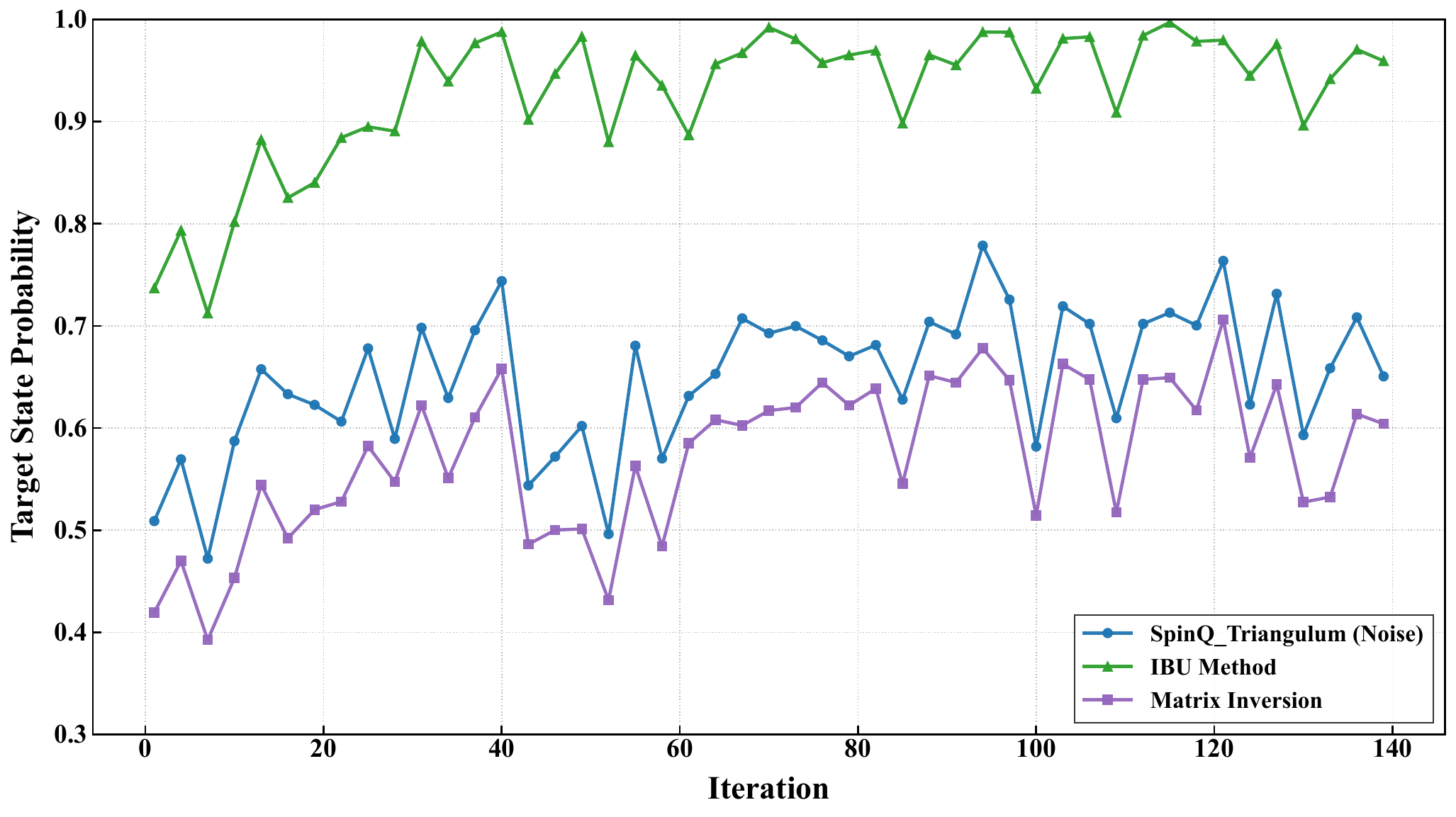}
        \caption{Target-state probability on Triangulum II.}
        \label{fig:triangulum_prob}
    \end{subfigure}
    \caption{Optimization trajectories on Triangulum II for the 5-city TSP instance. Panel (a) shows the loss values obtained from raw noisy measurements, iterative Bayesian unfolding (IBU), matrix-inversion-based mitigation, and the noise-free simulator. Panel (b) shows the corresponding target-state probabilities during optimization.}
    \label{fig:triangulum_hardware}
\end{figure}

The divide-and-conquer framework is further evaluated on quantum hardware using the two-qubit implementation described above. To examine the influence of experimental noise and measurement-error mitigation, four optimization trajectories are compared: raw hardware measurements, iterative Bayesian unfolding (IBU) mitigation~\cite{bauer2024ibu}, matrix-inversion-based measurement-error mitigation~\cite{bravyi2021mitigating}, and a noise-free simulator. The experiments are performed on SpinQ Gemini Pro and SpinQ Triangulum II NMR quantum computers. The corresponding results are shown in Figs.~\ref{fig:gemini_hardware} and~\ref{fig:triangulum_hardware}.

The hardware trajectories are broadly consistent with the noise-free reference, while showing deviations attributable to state-preparation errors, gate errors, decoherence, and measurement noise. For both devices, the computational basis state associated with the optimal tour remains among the dominant measured outcomes during the optimization, allowing the solution to be identified in this small instance.

Measurement-error mitigation changes both the reconstructed loss and the target-state probability. In these experiments, IBU yields trajectories closer to the noise-free reference than the raw data and is more stable than direct matrix inversion, particularly for the Triangulum II data. This observation is consistent with the greater sensitivity of matrix-inversion mitigation to statistical fluctuations and calibration uncertainty.

On SpinQ Gemini Pro, the raw hardware loss decreases from approximately \(101.81\) at the beginning of the optimization to \(98.75\) at the final iteration, while the target-state probability increases from \(0.714\) to \(0.829\). After IBU mitigation, the final loss is \(98.35\), and the final target-state probability is \(0.868\). Matrix-inversion mitigation yields a final loss of \(98.47\) and a final target-state probability of \(0.857\).

On SpinQ Triangulum II, the raw hardware loss fluctuates mainly around \(108\)--\(112\), whereas the IBU-mitigated loss is shifted toward the \(97\)--\(100\) range after the initial optimization stage. At the final iteration, IBU gives a loss of \(98.17\), compared with \(111.32\) for the raw hardware data and \(112.68\) for matrix inversion. The corresponding target-state probability is \(0.959\) under IBU, compared with \(0.651\) for the raw data and \(0.604\) for matrix inversion. The IBU-mitigated target-state probability remains close to or above \(0.9\) for much of the trajectory and reaches values above \(0.98\) at several iterations.

\begin{table}[htbp]
\centering
\caption{Final hardware results for the 5-city TSP instance. The table summarizes the final loss and target-state probability from the optimization trajectories in Figs.~\ref{fig:gemini_hardware} and~\ref{fig:triangulum_hardware}.}
\label{tab:hardware_summary}
\begin{tabular}{llcc}
\toprule
\textbf{Device} & \textbf{Measurement processing} & \textbf{Final loss} & \textbf{Final target-state probability} \\
\midrule
SpinQ Gemini Pro & Raw hardware & 98.75 & 0.829 \\
SpinQ Gemini Pro & IBU mitigation & 98.35 & 0.868 \\
SpinQ Gemini Pro & Matrix inversion & 98.47 & 0.857 \\
SpinQ Triangulum II & Raw hardware & 111.32 & 0.651 \\
SpinQ Triangulum II & IBU mitigation & 98.17 & 0.959 \\
SpinQ Triangulum II & Matrix inversion & 112.68 & 0.604 \\
\bottomrule
\end{tabular}
\end{table}

The different behavior of the two mitigation methods is plausibly related to their stability under noisy calibration and finite sampling. Direct matrix-inversion mitigation can amplify statistical fluctuations when the calibration matrix is imperfect or ill-conditioned. By contrast, IBU updates the reconstructed probability distribution iteratively while preserving physical probability constraints, which may explain its more stable behavior in the present data.

Overall, the hardware experiments show that the divide-and-conquer objective can be evaluated through two-qubit local circuits and can retain sufficient information to identify the optimal tour for the tested 5-city instance.

\section{Conclusion and Discussion}

This work presented a resource-efficient variational framework for the TSP under two hardware regimes. For devices that can support compact multi-register circuits, cyclic-symmetry reduction and binary city-label registers reduce the data-qubit scaling from \(O(n^2)\) in one-hot formulations to \(O(n\log n)\). The associated permutation-preserving ansatz performs variational search within the feasible tour subspace by applying parameterized register-wise SWAP operations, thereby avoiding repeated-city configurations from a valid initial ordering.

For devices with severe qubit limitations, the divide-and-conquer formulation decomposes the global variational state into smaller subsystem states and reconstructs the objective value from local measurements. This enables execution on two-qubit NMR processors, although it neglects inter-subsystem entanglement and introduces additional measurement and classical post-processing overhead.

Numerical results on 4-, 5-, and 6-city instances show that increasing the number of register-swap layers generally improves the success rate, with the best average success rates reaching \(100\%\), \(100\%\), and \(95.5\%\), respectively. Hardware experiments on SpinQ Gemini Pro and SpinQ Triangulum II further demonstrate that a 5-city instance can be implemented through two-qubit local circuits, where the optimal basis state remains identifiable under device noise. Iterative Bayesian unfolding provides the most stable recovery of the target-state probability among the tested measurement-error mitigation methods.

However, the present study remains limited to small-scale instances. The trainability and expressibility of the permutation-preserving ansatz for larger TSPs require further analysis, particularly in relation to barren-plateau effects in parameterized quantum circuits~\cite{mcclean2018barren}. In addition, the divide-and-conquer approximation reduces local qubit requirements by neglecting inter-subsystem entanglement, which may limit its accuracy for strongly correlated solution distributions and increase measurement overhead.

More broadly, resource efficiency in variational quantum algorithms can be pursued through several complementary routes. In addition to compact encodings and feasible-subspace circuit design, recent quantum-gradient-based optimization and full quantum variational eigensolver frameworks have reformulated the parameter-optimization stage to reduce resource overhead~\cite{Chen2024PureQuantumGradient}. The present framework contributes to this broader direction by focusing on qubit reduction, feasible-state preservation, and local objective reconstruction for permutation-constrained optimization.

Overall, these results show that compact encoding, feasible-subspace ansatz design, divide-and-conquer objective reconstruction, and measurement-error mitigation can be combined to make small structured optimization problems executable on strongly resource-limited quantum hardware. This provides an experimentally grounded route for studying permutation-constrained optimization in the NISQ regime.

\section{Acknowledgements}

This work was supported by the Shenzhen Science and Technology Program, China
(Grant No. JCYJ20241202123906009), and the National Natural Science Foundation
of China (Grant Nos. 12175002, 11705004, and 12381240288).

\section*{Data availability}

The generated TSP distance matrices, random seeds, optimal tours, and optimal tour lengths for the simulation benchmarks are provided in Supplementary Tables~S1--S3. Additional data supporting the hardware experiments, including raw measurement counts, readout calibration matrices, subsystem partitions, and processed probability distributions, are available from the corresponding author upon reasonable request.

\section*{Code availability}

The simulations were implemented using the SpinQit framework and a Torch-based simulator. Custom code used to generate TSP instances, construct the binary-register Hamiltonian, optimize the permutation-preserving ansatz, perform the divide-and-conquer reconstruction, and apply measurement-error mitigation will be made available from the corresponding author upon reasonable request.

\section*{Author contributions}

C.G. and C.Z. supervised the project. Y.L. developed the compact binary-register formulation, implemented the simulations, analyzed the numerical results, contributed to the algorithm design, and coordinated the hardware experiments. C.G. contributed to the algorithm design and coordinated the hardware experiments. Y.L. and C.G. interpreted the results and wrote the manuscript. All authors reviewed and approved the final manuscript.

\section*{Competing interests}

The authors declare no competing interests.

\section*{Additional information}
Correspondence and requests for materials should be addressed to C.G.
(cguo@spinq.cn). Yuefeng Lin may be contacted at y3lin@spinq.cn.

\bibliography{sample}

\end{document}